\documentstyle[sprocl]{article}

\input{psfig}
\def\be{\begin{equation}}
\def\ee{\end{equation}}
\def\bea{\begin{eqnarray}}
\def\eea{\end{eqnarray}}
\begin{document}

\title{IS QCD AT SMALL X A STRING THEORY?}

\author{R. Peschanski }

\address{
Service de Physique Th\'{e}orique
CEA-Saclay
F-91191 Gif-sur-Yvette
FRANCE}


\maketitle\abstracts{
Using the dipole picture describing the  $1/N_C$ limit of QCD at small $x$ and the conformal invariance properties of the BFKL kernel in transverse coordinate space, we show that the $%
1\!\!\rightarrow\!\!p$ dipole densities can be expressed in terms of dual
Shapiro-Virasoro amplitudes $B_{2p+2}$ and their generalization including non-zero conformal spins. We  discuss  the possibility of an effective closed string theory  of interacting QCD dipoles.
}

\section{Introduction}
 The  QCD ``hard Pomeron''  is understood as  the solution of perturbative QCD expansion at high energy ($W$) after resumming 
the leading $\left( \alpha \ln W^{2}\right)^n$ terms. It is known to obey 
the BFKL equation \cite{BFKL}. It has
recently attracted a lot of interest in relation with the experimental
results obtained at HERA for deep-inelastic scattering reactions at very low
value of $x\approx Q^{2}/W^{2},$ where $Q^{2}$ is the virtuality of the
photon probe $\gamma ^{*}$ and $W$ is, in this case, the c.o.m. energy of the $\gamma ^{*}$%
-proton system.\ Interestingly enough, the proton structure functions
increase with $W$ at fixed $Q$ in a way qualitatively compatible with the  prediction of the BFKL equation \cite{DESY}. However, the phenomenological discussion is still under way,
 since scattering of a ``hard'' probe on a proton is not a fully perturbative QCD process and, moreover, alternative explanations based on renormalization group evolution equations do exist \cite{DESY}. On the other hand, the phenomenological success \cite
{Proton} of parametrizations based on the BFKL evolution in the framework of the
QCD dipole model \cite{Muller} is quite encouraging for a further study of its properties.

Beyond these  phenomenological motivations, there exist interesting related
theoretical problems which we want to address in the present paper.\ In its
 2-dimensional version, the BFKL equation expresses \cite{BFKL} the leading-order resummation
result for  the elastic (off-mass-shell) gluon-gluon scattering amplitude in the
$2\!-\!d$ transverse plane. Alternatively one introduces \cite
{Lipatov,Muller}  the {\it space coordinate} variables via
2-dimensional Fourier transforms.
Explicit solutions of the BFKL equation  can be obtained \cite
{Lipatov,Navelet} using conformal invariance properties of the BFKL kernel expressed in the 
2-dimensional transverse coordinate space. Using these  symmetry properties in the sequel,
 we will address the problem of finding the vertices
for  processes
involving $2+2p$ external gluon legs where $\rho _{0}\rho
_{1}, $ $\rho _{a_0}\rho _{a_1},$...,$\rho _{p_0}\rho _{p_1}$ are their arbitrary
$2\!-\!d$ coordinates.
Our main result is the explicit solution (cf. formula (5)) of the $p$-uple dipole density distributions in the transverse coordinate plane, and their expression, (cf. formula (8))  as  integrands of dual Virasoro-Shapiro
amplitudes \cite {Shapiro-Virasoro,Frampton}. The expressions for arbitrary conformal spins (i.e. for all the  conformal components) are found to be analoguous to generalizations of  Shapiro-Virasoro amplitudes with excited states as external legs in a closed string theory.

\section{BFKL dynamics in the framework of the QCD dipole model}

Our starting point is the recently obtained result \cite {Navelet} that
 the $2+2$ amplitude is equal, up to kinematical factors, to  the
number density $n_{1}$ of dipoles  appearing in the $1/N_C$ QCD calculation of the wave-function of a massive quark-antiquark pair (onium) after an evolution  ``time''  $Y$ (such can be interpreted the total available rapidity space $Y$ in the BFKL equation written as a diffusion process \cite {bartels}). In the QCD-dipole picture \cite {Muller}%
, gluons are equivalent to  $q \bar q$ pairs (in the $N_{c}\!\rightarrow\!
\infty $ limit) which recombine into a collection of independent and colorless
dipoles. The elastic amplitude is thus obtained in the $N_{c}\!\rightarrow\!
\infty $ limit from the elementary
dipole-dipole amplitudes weighted by the dipole number densities of each
initial state obtained after evolution ``time'' $Y.$ Using conformal invariance properties of the BFKL
kernel, it turns out \cite{Lipatov,Muller} that:
\begin{eqnarray}
n_{1}\left( \rho _{0}\rho _{1};\rho _{0}^{\prime }\rho _{1}^{\prime
}|Y\right) &=&\int d\omega \ e^{\omega Y}\sum_{n\in  \cal Z} \int d\nu \ \frac{n_{1}^{n,\nu }\left( \rho _{0}\rho _{1};\rho
_{0}^{\prime }\rho _{1}^{\prime }\right) }{\omega -\omega \left(n,\nu
\right) }
\nonumber \\
&=&
\sum_{n\in  \cal Z} \int d\nu \ e^{\omega \left(n,\nu
\right) Y}\ n_{1}^{n,\nu }\left( \rho _{0}\rho _{1};\rho
_{0}^{\prime }\rho _{1}^{\prime }\right), 
\end{eqnarray}
\noindent where $n_{1}^{n,\nu }$  are the
components of the dipole density  expanded upon the conformally invariant
basis and
\begin{equation}
\omega \left( n,\nu \right) =\frac{2\alpha N_{c}}{\pi}\ \Re \left\{ \psi
\left( 1\right) -\psi \left(\frac {1+n}2 +i\nu \right) \right\}   
\end{equation}
\noindent is the value of the BFKL kernel in the (diagonal) conformal basis.
The corresponding eigenvectors are explicitly known \cite{Lipatov} to be:
\begin{equation}
E^{n,\nu }\left( \rho _{o\gamma },\rho _{1\gamma }\right) =\left( -1\right)
^{n}\left( \frac{\rho _{0\gamma }\rho _{1\gamma }}{\rho _{01}}\right)
^{\Delta_\gamma }\left( \frac{\bar{\rho} _{0\gamma }\bar\rho _{1\gamma }}{%
\bar{\rho} _{01}}\right)^{\widetilde{\Delta}_\gamma },  
\end{equation}
\noindent with $\rho _{ij}=\rho _{i}\!-\!\rho _{j} \left({\rm  resp.}\ \bar{\rho}
_{ij}=\bar{\rho _{i}}\!-\!\bar{\rho _{j}}\right) $ are the
holomorphic (resp. antiholomorphic) components in the 2-$d$ transverse plane
represented by the complex plane  ${ \cal C}$ and $\Delta_\gamma =\frac {n-1}2-i\nu$, $ \widetilde{\Delta}_\gamma %
=-\frac {n+1}2-i\nu,$ ($n\!\in \! { \cal Z}$, ${\nu }\!\in  \!{ \cal R}$), are the
quantum numbers defining the appropriate unitary representations \cite{Lipatov} of the global conformal group $SL(2,{ \cal C})$. Indeed the BFKL solution (i.e.,
also, the QCD dipole solution)
 is given by 
\begin{equation}
n_{1}^{n,\nu }\left( \rho _{0}\rho _{1};\rho
_{0}^{\prime }\rho _{1}^{\prime }\right) =\frac{%
\nu^{2}+n^{2}/4}{\vert\rho
_{0^{\prime }1^{\prime} }\vert ^2 \ \pi ^{4}/2} \ \int_{{ \cal R}^{2}}d^{2}\rho _{\gamma }\ \bar{E}
^{n,\nu}\left( \rho _{0\gamma }^{\prime },\rho _{1\gamma }^{\prime }\right)
E^{n,\nu}\left( \rho _{0\gamma ,}\rho _{1\gamma }\right) ,  \end{equation}
which can be explicitly calculated in terms of hypergeometric functions \cite{L.N.,Navelet}.

\section{Multiple-dipole vertices are dual string amplitudes}

 In order to generalize these investigations to an arbitrary number of
gluons, we shall use the QCD dipole formalism allowing to express the probability of finding $p$ dipoles in the wave-function of an initial one, i.e. the 
$p$-uple dipole density after an evolution $Y,$ 
$n_p\left( \left. \rho _{0}\rho _{1};\rho _{a_0}\rho _{a_1},\rho _{b_0}\rho
_{b_1},...,\rho _{p_0}\rho _{p_1}\right| Y\right) .$ $n_p$ is the solution of an integral equation which has been proposed in Ref. \cite{A. Muller},
and approximate solutions have been worked out and applied to problems like the
double and triple QCD Pomeron coupling \cite{A. Muller}, dipole production \cite{Bialas1}, hard diffraction \cite{Bialas} and, more
generally, to the unitarization problem \cite{A. Muller}. In particular, Monte-Carlo
simulations of the unitarization series based on a numerical resolution of the 
$n_p$ integral equations have been performed \cite{muller}. 

However a general expression for the solution of
these equations and a physical interpretation of its properties were still lacking. It is the purpose of our work to provide
such a solution, which is intimately related, as we shall see, to dual 
string amplitudes emerging from the QCD dipole picture.

We now  consider  the p-uple distribution of dipoles $n_{p}^{n,\nu } .$  One writes:
\begin{eqnarray}
&& n_{p}^{n,\nu }\left( \left. \rho _{\gamma };\rho _{a_0}\rho _{a_1},...,
\rho _{p_0}\rho _{p_1}\right| {\omega }\right) =\frac{1}{%
2a(n,\nu) \left( \omega -\omega {\left( n,\nu \right) }\right) 
}\frac{1}{\left| \rho _{a}... \rho _{p}\right| }\ \times \nonumber \\
&& \!\!\!\!\!\!\!\!\sum_{n_{a},...,n_{p}}\int \frac{d\nu _{a}...d\nu
_{p}}{a{\left( n_{a},\nu _{a}\right) }...a{\left( n_{p},\nu _p\right) }}\ \frac{1}{\omega {\left( n_{a},\nu
_{a}\right) }+...+\omega {\left(
n_{p},\nu _{p}\right) }-\omega}   \nonumber \\
 \!\!\!\!\!\!\!\! &&\int d^{2}\rho _{\alpha }...d^{2}\rho
_{\pi }
\;\bar{E}^{n_{a},\nu _{a}}{\left( \rho _{a_{0}\alpha},\rho _{a_{1}\alpha}
\right)}
\ ...
\;\bar{E}^{n_{p},\nu _{p}}{\left( \rho _{p_{0}\pi},\rho _{p_{1}\pi}
\right)}\times
  \nonumber \\
\!\!\!\!\!\!\!\! &&\footnotesize{\int \frac{d^{2}\rho _{0}...d^{2}\rho _{p}%
}{\left| \rho _{01}\ \rho _{12} ... \rho _{p0}\right| ^{2}}\ E^{n,\nu }\left( \rho
_{0\gamma },\rho _{1\gamma }\right) E^{n_{a},\nu _{a}}{\left( \rho _{1\alpha},\rho _{2\alpha}
\right)}
...E^{n_{p},\nu _{p}}\left( \rho _{p\pi},\rho _{0\pi}\right)}  
\end{eqnarray}
\noindent where we have used the fact that the probability kernel of finding $p$ dipoles at ``time'' $Y$ can be expressed by   iteration of the single BFKL kernel, namely:
\begin{equation}
\left|\frac {\rho _{01}}{ \rho _{02}\ \rho _{12}}\right|^2\times
\left|\frac {\rho _{02}}{ \rho _{23}\ ... \rho _{(p\!-\!1)p}\rho _{p0}}\right|^2
=
\left|\frac {\rho _{01}}{ \rho _{12}\ ... \rho _{(p\!-\!1)p}\rho _{p0} }\right|^2.
\end{equation}
Hence, the overall kernel is expressed in a symmetric way as a function of the coordinates of the perimeter designed by the coordinates ${\rho _{0}},{ \rho _{1}\ ... \rho _{p\!-\!1}}, \rho _{p}$  independently of the possible intermediate steps   (cf. the ${\rho _{0}}{ \rho _{2}}$ segment).

Let us consider the last term, i.e. the vertex part $V_{\gamma...\pi},$ of formula (5).  
Inserting the definitions (3) in the expression of $V_{\gamma...\pi},$ one gets: \begin{equation}
V_{\gamma...\pi}={\int_{{ \cal C}^{\ p+1}} }\frac{d^{2}\rho
_{0}...d^{2}\rho _{p}}{\left| \rho _{01}...\rho
_{p0}\right| ^{2}}\ \times\left( \frac{\rho _{0\gamma }\rho _{1\gamma }}{\rho
_{01}}\right) ^{\Delta_\gamma }...\left( \frac{\rho _{p\pi }\rho _{0\pi }}{%
\rho _{p0}}\right) ^{\Delta_\pi }\times \left\{\rho \Rightarrow \bar\rho, \Delta \Rightarrow \tilde \Delta \right\}.  
\end{equation}
Our observation is that \medskip $V_{\gamma...\pi }$ can be
expressed as the integrand of a Shapiro- Virasoro amplitude $B_{2p+2}.$
One may write
\begin{eqnarray}
B_{2p+2}={\int_{{ \cal C}^{ p+1}} }\frac {d^{2}\rho _{\alpha }...d^{2}\rho
_{\pi }}{ { \cal V}_3 }\ 
\prod\limits_{\eta_i<\eta_j}^{p+1}
\ \rho _{\eta_i\eta_j}^{-\Delta_{\eta_i\eta_j}}
\ \bar {\rho} _{\eta_i\eta_j}^{-{\tilde \Delta_{\eta_i\eta_j}}}\ V_{\gamma...\pi}=\nonumber \\
={\int_{{\cal C}^{ 2+2p}} } \frac {d^{2}\rho _{\alpha }...d^{2}\rho
_{\pi }}{ { \cal V}_3 }\ d^{2}\rho
_{0}...d^{2}\rho _{p}\ \left( \rho _{0\gamma }\rho _{1\gamma }\right)^{\Delta_\gamma }
\left( \rho _{1\alpha }\rho _{2\alpha }\right)^{\Delta_\alpha }...
\left( \rho _{p\!\!-\!\!1\pi }\rho _{p\pi }\right)^{\Delta_\pi }
\nonumber \\ \times \prod\limits_{\eta_i<\eta_j}^{p+1}
\ \rho _{\eta_i\eta_j}^{-\Delta_{\eta_i\eta_j}}
\
\left(
{\rho
_{01}}\right) ^{-1-\Delta_\gamma }...\left( \rho _{p0}\right)
^{-1-\Delta_\pi }\times \left\{\rho \Rightarrow \bar\rho, \Delta \Rightarrow \tilde \Delta \right\},
\end{eqnarray}
where the conformal dimensions $\Delta_{\eta_i\eta_j}, (\eta_i, \eta_j \in (\alpha...\pi)),$ are fixed by global  $SL\left( 2,{ \cal C}\right)$ invariance \cite {Polyakov,difrancesco}.
The factor ${ \cal V}_3^{-1}$ formally corresponds  to the ``division'' by the volume of the group $SL\left( 2,{ \cal C}\right) .$ In practice \cite{Koba-Nielsen} one may arbitrarily  fix three coordinates $\rho _{\eta_1}, \rho _{\eta_2}, \rho _{\eta_3},$
and write
\begin{equation}
{ \cal V}_3 \equiv \frac{d^{2}\rho _{\eta_1}d^{2}\rho _{\eta_2}%
d^{2}\rho _{\eta_3}}{\left| \rho _{\eta_1}\rho _{\eta_2}\rho _{\eta_3}
\right| ^{2}}.
\end{equation}

Eqn.(14) is a particular realization of the Shapiro Virasoro amplitudes obtained as  bosonic string tree amplitudes. Indeed, 
by definition \cite{Shapiro-Virasoro,Frampton}, one writes:
\begin{equation}
B_{N}=\int \prod\limits_{j=1}^{N}\frac {d^2\rho _{j}}{ { \cal V}_3 }\ \ 
\prod\limits_{i<j}^{N}
\ \rho _{ij}^{-p_{ij}}
\ \bar {\rho} _{ij}^{-{\tilde p}_{ij}},  
\end{equation}
where the powers $p_{ij}$ and ${\tilde p}_{ij}$ obey a set of $SL\left( 2,{ \cal C}\right) $ constraints \cite{Frampton}. These constraints are automatically satisfied in our case, since the global invariance is respected at every step.
The powers $p_{ij}$  appearing in (8)  are known combinations of conformal dimensions $\Delta$ and  $\tilde \Delta$ corresponding to  particular realizations of the  $SL\left( 2,{ \cal C}\right) $ constraints.

\section{A closed string theory for  QCD at low x?}

 The Shapiro-Virasoro amplitudes which are obtained for  $(1\!\rightarrow \! p)$ dipole distributions lead naturally to  the question of
a closed-string interpretation of the high-energy limit of perturbative QCD. Indeed, these amplitudes appear in the  context of a closed string moving in a Minkowskian $(1,d\!-\!1)$ target space \cite {Frampton}. More generally, such amplitudes appear as a consequence of vertex operator constructions in conformal field theories \cite {difrancesco} and are related to the existence of an (anomalous) infinite-dimensional algebra associated with local conformal invariance, namely the Virasoro algebra. Moreover in the case of a critical target-space dimension $d_c,$ the Fock space on which the quantum string theory is realized is spanned by positive normed states (no ghosts) with full reparametrization invariance. This connection has both a practical and conceptual interest for QCD calculations. First, the many and much explored mathematical properties of dual amplitudes may lead to a simplification of QCD dipole computations for given processes, e.g. ``hard'' diffraction, multi-Po!
!
meron contributions, etc.. Second, there is a possibility of building an effective theory of QCD in the high-enrgy limit, which could be based on a string theory (instead of a field theory). This would allow the computation of string loop contributions and thus induce an effective theory of interacting QCD Pomerons. 

However, the variables which appear as conformal exponents $p_{ij}$
of the integrands are not directly expressed as scalar products of momenta in a 
Minkowskian $(1,d\!-\!1)$ target space. They are complex numbers
 obeying constraints which are not directly expressed as on-mass shell and momentum conservation constraints as for the closed string \cite {Frampton}. Even if such a target-space interpretation is possible, an analytic continuation in the imaginary direction (implied by the quantum numbers of the conformal eigenvectors (3)) is to be performed. It is thus useful to pass in review the  properties of Shapiro-Virasoro amplitudes in this context and to see which are those to be completed for a full closed string theory to be valid.

\quad i)\quad {\bf Duality}

A first consequence of the solutions (5),(8) is that duality properties exist in the dipole formulation of QCD vertices. Indeed, by construction, Shapiro- Virasoro amplitudes are meromorphic and forbid the existence of multiple pole singularities coming from  dual channels. For instance,
the $(\rho_0,\rho_2)$ and $(\rho_1,\rho_i),$ $ i=3...p$  channels appearing in (5), (6) cannot   bring  simultaneous singularities to the amplitude. As usual in dual theories, there exists intricate relations between
different ways of describing the amplitudes depending on the series of 
pole contributions which are choosen for their expansion. An interesting first example of such a duality property has been indeed provided by the equivalence of the ``t-channel'' BFKL elastic 4-gluon amplitude with the ``s-channel''
QCD dipole description of the same amplitude \cite {Navelet}. Further applications of this fruitful concept are expected from our results.

\quad ii)\quad {\bf Non-zero conformal spins}
 
As a practical consequence of our identification of the
multiple-dipole vertices with  integrands of standard Shapiro-Virasoro amplitudes in the case of zero conformal spins, one may use some  tools \cite{kawai} which are developed in the string theoretical formalism
to generalize our investigations to the case of general (integer or
half-integer) conformal spins as follows; One can consider in general amplitudes of the form:
\begin{equation}
B_{N}=\int \prod\limits_{i=1}^{N}d\rho _{i}\left[ \frac{d\rho
_{\alpha }d\rho _{\beta }d\rho _{\delta }}{\left| \rho _{\alpha \beta }\rho
_{\beta \delta }\rho _{\delta \alpha }\right| ^{2}}\right]
^{-1}\prod\limits_{i<j}^N
\ \rho _{ij}^{-p_{ij} + \frac {n_{ij}}2}
\bar {\rho} _{ij}^{-p_{ij} + \frac {{\tilde n}_{ij}}2},  \end{equation}
\noindent where $n_{ij},\ {\tilde n}_{ij} $ are integers.
Interestingly enough, in the framework of string theory,  this corresponds to consider external excited states
of the bosonic string. Moreover, the same techniques  allow to connect
closed string to open string tree amplitudes which may allow to extend to the
multiple-vertex calculations the conformal-block structure initially identified in the BFKL 4-point amplitudes \cite{Navelet}.

\quad iii)\quad {\bf Extended conformal symmetry and Virasoro algebra} 

In the original paper of Ref. \cite {Lipatov}, it has been noticed that it was not straightforward to extend the (already beautiful) global conformal symmetry $SL(2,{ \cal C})$ to the infinitely dimensional conformal group in 2 dimensions. In other words, only the 6 generators of the Virasoro algebra: ${ \cal L}_{-1},$ ${ \cal L}_{1},$ ${ \cal L}_{0}$ (3 holomorphic) and their 3 non-holomorphic counterparts, were expected to generate the symmetry algebra of the BFKL kernel. The results we obtain indicate that the algebra can probably be extended to the infinite series
of locally conformal generators ${ \cal L}_{n}, n \in { \cal Z}$, i.e. the whole Virasoro algebra, at least in the QCD dipole representation. as usual, the symmetry is expected to be anomalous due to the possibility of  a central charge \cite {Frampton,difrancesco} (conformal anomaly) at the quantum level of consistency. This issue will depend on the interpretation of a suitable $p-$independent target-space representation of the exponents $p_{ij}.$  For instance, if an embedding  $p_{ij}\!\rightarrow \!p_i\cdot p_j$ in a Minkowskian   $(1,d\!-\!1)$ space is allowed, this will determine the central charge to be related to $d$ and the critical dimension to be $d_c = 26,$ by compensation of the ghost contribution due to  reparametrization symmetry \cite {Frampton}. This interesting issue   certainly deserves more study. 

\section*{Acknowledgments}
\vspace{-4mm}
We want to  thank Andrzej Bialas and Henri Navelet for a fruitful collaboration on the QCD theory of dipoles which initiated the present work.  Christophe Royon and Samuel Wallon are acknowledged for stimulating discussions.

\end{document}